\documentclass[prd,twocolumn,superscriptaddress,amsmath,amssymb,nofootinbib]{revtex4-1}

\usepackage{graphicx}

\begin{document}

\title{Towards the Uniqueness of Lifshitz Black Holes and Solitons in New Massive Gravity}

\author{Eloy Ay\'on--Beato}
\email{ayon-beato-at-fis.cinvestav.mx}
\affiliation{Departamento~de~F\'{\i}sica,~CINVESTAV--IPN,%
~Apdo.~Postal~14--740,~07000,~M\'exico~D.F.,~M\'exico}
\affiliation{Instituto~de~Ciencias~F\'isicas~y~Matem\'aticas,%
~Universidad~Austral~de~Chile,~Casilla~567,~Valdivia,~Chile}

\author{Mokhtar Hassa\"{\i}ne}
\email{hassaine-at-inst-mat.utalca.cl}
\affiliation{Instituto~de~Matem\'atica~y~F\'{\i}sica,%
~Universidad~de~Talca,~Casilla~747,~Talca,~Chile.}

\author{Mar\'ia Montserrat Ju\'arez-Aubry}
\email{mjuarez-at-fis.cinvestav.mx}
\affiliation{Departamento~de~F\'{\i}sica,~CINVESTAV--IPN,%
~Apdo.~Postal~14--740,~07000,~M\'exico~D.F.,~M\'exico}
\affiliation{Instituto~de~Ciencias~F\'isicas~y~Matem\'aticas,%
~Universidad~Austral~de~Chile,~Casilla~567,~Valdivia,~Chile}

\begin{abstract}
We prove that the $z=1$ and $z=3$ Lifshitz black hole solutions of New Massive Gravity in three dimensions are the only static axisymmetric solutions that can be cast in a Kerr-Schild form with a seed metric given by the Lifshitz spacetime. Correspondingly, we study the issue of uniqueness of Lifshitz solitons when considering an ansatz involving a single metric function. We show this problem can be mapped to the previous one and that the $z=1$ and $z=1/3$ Lifshitz soliton solutions are the only ones within this class. Finally, our approach suggests for the first time an explanation as to what is special about those particular values of the dynamical critical exponent $z$ at finite temperature.
\end{abstract}

\maketitle

\section{Introduction}

In the last five years, there has been an intense activity in order to export the ideas of the anti-de Sitter/Conformal Field Theory (AdS/CFT) correspondence \cite{Maldacena:1997re} to non-relativistic physics. This is motivated by the fact that some models that describe this kind of phenomena enjoy the non-relativistic version of the conformal symmetries: the so-called  Schr\"odinger group \cite{Jackiw:1972cb,Niederer:1972zz}. In the spirit of the AdS/CFT correspondence, the gravity dual metric has been identified as the spacetime whose isometry group coincides with the Schr\"odinger group, see Refs.~\cite{Son:2008ye} and \cite{Balasubramanian:2008dm}. Of particular importance is also the construction of Kachru \emph{et al.} \cite{Kachru:2008yh} which proposed a new gravitational background dual to scale invariant Lifshitz fixed points without Galilean boost invariance. In this case the gravity backgrounds are commonly called Lifshitz spacetimes, and its simplest realization occurs in three dimensions as follows:
\begin{equation}\label{eq:Lifshitz}
ds_L^2=-\frac{r^{2z}}{l^{2z}}dt^2+\frac{l^2}{r^2}dr^2+\frac{r^2}{l^2}dx^2.
\end{equation}
These spacetimes admit as their main feature an anisotropic scaling symmetry
\begin{equation}\label{eq:scaling}
t \mapsto \lambda^z t, \qquad
x \mapsto \lambda x, \qquad
r \mapsto {\lambda}^{-1} r ,
\end{equation}
with dynamical critical exponent $z>0$ (the isotropic case $z=1$ corresponds to AdS) as part of their isometry group. Black hole solutions whose asymptotic behaviors reproduce the Lifshitz metric are of particular importance for a finite-temperature holographic description of non-relativistic systems and are commonly known as Lifshitz black holes. However, in contrast with the AdS case, Lifshitz spacetimes are not solutions of general relativity and require the introduction of matter sources or to consider square-curvature corrections to standard gravity \cite{Adams:2008zk,Cai:2009ac, AyonBeato:2009nh,AyonBeato:2010tm}. The simplest example of these theories is the so-called new massive gravity (NMG) \cite{Bergshoeff:2009hq} in three dimensions, which has been intensively studied in the last five years. This theory presents the advantage of being at the linearized level equivalent to the unitary Fierz-Pauli theory for free massive spin-2 gravitons. Up to now, there are only two known static black hole solutions of this theory both characterized asymptotically by a specific dynamical exponent $z$. The first one occurs for $z=1$ and it is nothing but an asymptotically AdS black hole solution derived in \cite{Bergshoeff:2009aq,Oliva:2009ip} while the second one corresponds to a proper Lifshitz black hole solution with a dynamical exponent $z=3$ and found in \cite{AyonBeato:2009nh}.  These black holes have Lifshitz soliton counterparts obtained in \cite{Oliva:2009ip}, \cite{Gonzalez:2011nz} by operating a double Wick rotation in the Killing $(t,x)$-plane which are characterized by dynamical exponents $z=1$ and $z=1/3$, respectively. As shown in \cite{Gonzalez:2011nz}, this is due to the fact that the two-dimensional Lifshitz algebras with dynamical exponents $z$ and $z^{-1}$ are isomorphic.

Unlike in General Relativity \cite{Chrusciel:2012jk}, it is extremely difficult to establish uniqueness results in higher-order gravity theories, which are by default plagued of instabilities. The unitarity of NMG opens a window of opportunity, however, since there exist at least two different static black hole solutions, such uniqueness theorem cannot be established. Therefore, it is important to understand what is the spectrum of black hole configurations admitted by this theory. A reasonable starting point is the fact that most of the black hole solutions that we know can be expressed as a Kerr-Schild transformation from the spacetime to which they approach asymptotically.\footnote{One of the few exceptions is the five-dimensional black ring \cite{Emparan:2001wn}.} It is well-known that the original Kerr-Schild ansatz \cite{KS}, as well as its generalizations, is a powerful tool for searching stationary black hole solutions of Einstein equations both in vacuum and in presence of matter sources, see e.g.~\cite{Dereli:1986cm} and more recently \cite{Gibbons:2004uw} for higher dimensions. Just for completeness, we also mention the existence of examples for which the Kerr-Schild ansatz may be efficient even with a seed metric that does not solve the field equations, see e.g.~\cite{Anabalon:2009kq}, \cite{Ett:2010by}. The advantage of the Kerr-Schild ansatz usually lies in the fact that it linearizes a certain combination of curvature components and since standard gravity is linear in curvature, the Kerr-Schild ansatz consequently linearizes Einstein equations. This is no longer the case for higher-curvature theories. However, part of the dynamics is still linearized for theories with squared-curvature corrections as NMG. This property will be useful in order to achieve the main purpose of this work: to understand what are the asymptotically Lifshitz static, axisymmetric black holes admitted by NMG that can be written as a Kerr-Schild transformation from the Lifshitz spacetime (\ref{eq:Lifshitz}). In fact, we will show that the dynamics of NMG restricts those configurations via two constraints: a linear one, due to the Kerr-Schild ansatz, and a non-linear one capturing the natural higher-curvature character of the theory. After a thorough analysis, we conclude that the struggle between both behaviors is only consistently balanced for the values $z=1$ and $z=3$ of the critical exponent, yielding to the previously mentioned black hole solutions. In other words, the solution space is a constrained system within a linear space and many of the linear directions end up being inconsistent with the nonlinear constraint. Additionally, these precise exponents play a special role since they are the only ones for which the characteristic polynomial determining the linear directions presents multiplicities, i.e.\ they are the boundaries between different behaviors of the linear spaces. Even if this is only a partial uniqueness statement, it is actually one of the very few uniqueness results found in high-order gravity theories. A similar analysis is carried out for the Lifshitz soliton solutions of NMG, which cannot be cast in Kerr-Schild form but likewise allow a representation in terms of a metric ansatz involving a single function. Remarkably, the soliton field equations can be mapped to the black hole ones. This allows us to find that the only soliton solutions within this class are those obtained from the now unique black hole solutions operating a double Wick rotation, and yielding to the Lifshitz solitons with dynamical exponents $z=1$ and $z=1/3$.

Let us now present the different ingredients in order for the paper to be self-consistent. The action of new massive gravity is given by
\begin{equation}\label{eq:S}
S[g_{\mu\nu}]=\frac{1}{2\kappa}\int{d^3x}\sqrt{-g}
\left( R - 2\lambda - \frac{1}{m^2}K \right),
\end{equation}
where $R$, $R_{\mu\nu}$ and $\lambda$ stand respectively for the Ricci scalar, the Ricci tensor and the cosmological constant, while the unitary correction is defined as $K=R_{\mu\nu}R^{\mu\nu}-(3/8)R^2$. The associated field equations read
\begin{subequations}\label{eq:NMG}
\begin{equation}
E_{\mu\nu}\equiv R_{\mu\nu}-\frac{1}{2}Rg_{\mu\nu}+\lambda{g}_{\mu\nu}-\frac{1}{2m^2}K_{\mu\nu}=0,
\end{equation}
where the higher-order contribution
\begin{eqnarray}
K_{\mu\nu}&=&2\Box R_{\mu\nu} -\frac{1}{2}
\left( g_{\mu\nu}\Box R+\nabla_{\mu}\nabla_{\nu}R \right)
-8R_{\mu\alpha}R^{\alpha}_{~\nu} \nonumber\\
&& {}+\frac{9}{2} RR_{\mu\nu}
+g_{\mu\nu}\left( -\frac{13}{8}R^2+3R^{\alpha\beta}R_{\alpha\beta} \right),
\end{eqnarray}
\end{subequations}
has second order trace $g^{\mu\nu}K_{\mu\nu}=K$.

As was originally shown in Ref.~\cite{AyonBeato:2009nh}, the field equations of NMG (\ref{eq:NMG}) allow Lifshitz spacetimes (\ref{eq:Lifshitz}) as solutions for a generic value of the dynamical exponent $z$, by means of the following suitable parameterizations of the cosmological constant and the mass
\begin{subequations}\label{eq:ccz<>1}
\begin{eqnarray}
\lambda &=&-\frac{z^2+z+1}{2l^2},\label{eq:lambda}\\
    m^2 &=&-\frac{z^2-3z+1}{2l^2},\label{eq:mp}
\end{eqnarray}
\end{subequations}
where the values of $z$ giving a vanishing value of the mass are forbidden. These parameterizations are valid in general except
for the AdS case $z=1$, for which only the following constraint is satisfied
\begin{equation}\label{eq:lambda2mz=1}
\lambda = -\frac1{l^2}\left(1+\frac1{4l^2m^2}\right).
\end{equation}
The different static black hole and soliton solutions will be presented at the beginning of each Section before proving their uniqueness in the following subsections. Conclusions and comments on possible applications to other contexts of the arguments introduced in this work are provided at the end.

\section{Kerr-Schild Ansatz for Lifshitz black holes of new massive gravity}

The first static black hole solution found for NMG was the $z=1$ AdS black hole, independently obtained by Bergshoeff \emph{et al.} \cite{Bergshoeff:2009aq} and Oliva \emph{et al.} \cite{Oliva:2009ip}. The line element is given by
\begin{subequations}\label{eq:z=1bh}
\begin{eqnarray}
ds^2&=&-\frac{r^2}{l^2}\left(1-\mu\frac{l^2}{r^2}-b\frac{l}{r}\right)dt^2\nonumber\\
&&   {}+\frac{l^2}{r^2}\left(1-\mu\frac{l^2}{r^2}-b\frac{l}{r}\right)^{-1}dr^2
+\frac{r^2}{l^2}dx^2,
\end{eqnarray}
and the solution occurs for
\begin{equation}\label{eq:ccz=1}
\lambda=m^2=-\frac{1}{2l^2}.
\end{equation}
\end{subequations}
The extra constant $b$ represents a gravitational ``hair'' encoding the higher-order contribution. Note that for $b=0$, the solution becomes the Ba\~nados-Teitelboim-Zanelli (BTZ) black hole \cite{Banados:1992wn} which in general obeys the weaker constraint (\ref{eq:lambda2mz=1}).

The other static black hole solution known for NMG is the $z=3$ Lifshitz black hole found in \cite{AyonBeato:2009nh}, and its line element reads
\begin{subequations}\label{eq:z=3bh}
\begin{eqnarray}
ds^2&=&-\frac{r^6}{l^6}\left(1-M\frac{l^2}{r^2}\right)dt^2\nonumber\\
   &&{}+\frac{l^2}{r^2}\left(1-M\frac{l^2}{r^2}\right)^{-1}dr^2
   +\frac{r^2}{l^2}dx^2,
\end{eqnarray}
with coupling constants given by
\begin{equation}\label{eq:ccz=3}
\frac{\lambda}{13}=m^2=-\frac{1}{2l^2}.
\end{equation}
\end{subequations}
In both cases the identification $x=x+2{\pi}l$ is understood.

As will be appreciated below, both black holes allow a Kerr-Schild representation. In its most general form, a Kerr-Schild transformation permits the construction of a spacetime that asymptotes a starting one using a scalar function and a null
geodesic vector field of the original metric. In our problem, the relevant Kerr-Schild ansatz for a stationary axisymmetric spacetime is the following
\begin{equation}\label{eq:KS}
ds^2=ds^2_L + \bar{H}(r)k \otimes k,
\end{equation}
where $ds_L^2$ corresponds to the asymptotic spacetime, namely the Lifshitz metric (\ref{eq:Lifshitz}). It is straightforward to show that the two more general families of null geodesics compatible with the translation symmetries of the Lifshitz spacetime (\ref{eq:Lifshitz}), i.e.\ being stationary and axisymmetric are given by
\begin{equation}
k_{\pm}(a)=dt-adx\pm\frac{l^2}{r^2}\left(\frac{l^{2(z-1)}}{r^{2(z-1)}}
-a^2\right)^{1/2}dr.
\end{equation}
However, if we were to preserve the contribution along the $x$ direction, we would lose the asymptotically Lifshitz behavior in the Kerr-Schild transformation (\ref{eq:KS}), except for $z=1$. This is the reason why we keep our analysis within the static setup fixing the angular momentum $a$ to zero and choose as null geodesic vector $k=k_{+}(0)$. We found convenient to redefine the scalar function as $\bar{H}(r)=(r^{2z}/l^{2z}) H(r)$. Moreover, in order to get a manifestly static form of the transformed metric
we have to perform a Boyer-Lindquist like time redefinition
\begin{equation}
t \mapsto t + \int{\frac{l^{z+1}}{r^{z+1}}\frac{H(r)dr}{1-H(r)}},
\end{equation}
allowing the transformed metric to become
\begin{equation}\label{eq:Lifansatz}
ds^2=-\frac{r^{2z}}{l^{2z}}(1-H)dt^2
     +\frac{l^2}{r^2}\frac{dr^2}{1-H}
     +\frac{r^2}{l^2}dx^2.
\end{equation}
This is the most general static axisymmetric spacetime that can be cast in a Kerr-Schild form with a seed metric given by the Lifshitz spacetime.
Here the boundary condition warranting the Lifshitz asymptotic is simply $H(\infty)=0$. Notice that both black holes (\ref{eq:z=1bh}) and (\ref{eq:z=3bh}) are exactly represented in the above way for convenient elections of the function $H$ for each exponent.

Starting with the ansatz (\ref{eq:Lifansatz}), the field equations of NMG
(\ref{eq:NMG}) are given by the three non-trivial diagonal equations
$E_{t}^{~t}=0$, $E_{r}^{~r}=0$ and $E_{x}^{~x}=0$. However,
these three equations in spite of being linearly independent,
are not differentially independent, since we have the Bianchi-like relation
\begin{eqnarray}
(E_{r}^{~r})'&=&
\frac12\left[\ln\left(\frac{r^{2z}}{l^{2z}}(1-H)\right)\right]'
(E_t^{~t}-E_r^{~r})\nonumber\\
&&{} +\frac{E_{x}^{~x}-E_r^{~r}}{r}.
\end{eqnarray}
As a consequence, it is sufficient to consider linear
combinations of the equations $E_{t}^{~t}=0$ and $E_{r}^{~r}=0$
to work with. In our case, we consider the following combinations
\begin{widetext}
\begin{subequations}
\begin{eqnarray}
\frac{4m^2l^4} {1-H}(E_{r}^{~r}-E_{t}^{~t})&=& r^4H''''+2(z+4)r^3H'''-(z^2-17z-8)r^2H''-2(z+2)(z^2-5z+2)rH'\nonumber\\
&&{}-2(z-1)(z^2-3z+1)H+2(z-1)(2l^2m^2+z^2-3z+1)=0 \label{eq:elf},\\\nonumber\\
-16m^2l^4E_{t}^{~t}&=&
4(1-H)r^4H''''-2\left[rH'-2(3z+7)(1-H)\right]r^3H'''+(r^2H'')^2\frac{}{}\nonumber\\
&&{}-2\left[3rH'-2(z^2+17z+6)(1-H)\right]r^2H''+(5z^2-14z+5)(rH')^2\frac{}{}\nonumber\\
&&{}-4\left[(3z^3-13z^2-8z+5)(1-H)-2l^2m^2\right]rH'\frac{}{}\nonumber\\
&&{}+4(z^2-3z+1)(z^2+z-1)(1-H)^2-16l^2m^2(1-H+l^2\lambda)=0.
\quad~\label{eq:eln}
\end{eqnarray}
\end{subequations}
\end{widetext}

Particularly interesting is the linearity of the combination
(\ref{eq:elf}) which results as a straightforward consequence of the Kerr-Schild ansatz. In this case, the linearity is determined by a fourth-order Euler differential
equation. This equation presents an inhomogeneity whose constant particular solution is incompatible with the Lifshitz boundary condition $H(\infty)=0$ unless the inhomogeneity vanishes, which occurs for $z=1$ or for the parametrization (\ref{eq:mp}) if $z\neq1$ (in this equation it is clear why the values of $z$ giving a vanishing of the mass are forbidden). The linear space of solutions of the resulting homogeneous equation is spanned by power-laws $(l/r)^\alpha$, where the powers $\alpha$ satisfy the following characteristic polynomial
\begin{eqnarray}\label{auxiliar}
&&\alpha^4-2(z+1)\alpha^3-(z^2-11z+5)\alpha^2\nonumber\\
&&{}+(z+1)(2z^2-9z+6)\alpha-2(z-1)(z^2-3z+1)=0.\nonumber\\
\end{eqnarray}%
\begin{figure}[h]
\centering
\includegraphics[scale=1]{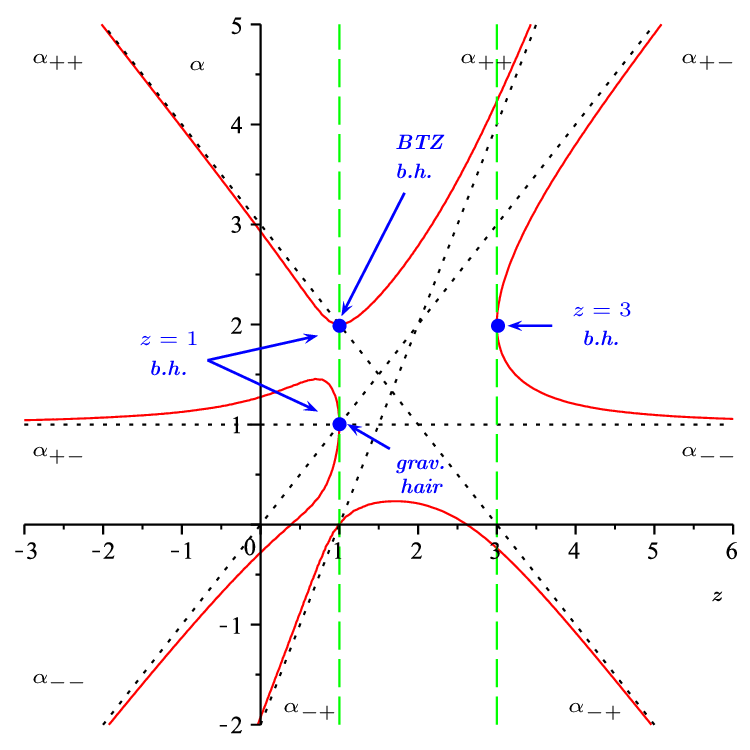}
\caption{The space of black hole solutions of NMG is encoded in this graph. The real roots $\alpha_{\pm\pm}$ of the characteristic polynomial (\ref{auxiliar}) are represented as red solid lines [see their definition in (26a)]. Notice that while most values of $z$ correspond to four roots of $\alpha$, multiplicity arises in $z=1$ and $z=3$ (both in green dashed lines) as roots $\alpha=1$ and $\alpha=2$ respectively are degenerate. Also, in the interval $1<z<3$ two of the roots
are actually complex conjugate and only the two real solutions appear in the figure. The asymptotes of the characteristic polynomial (\ref{auxiliar}) are represented by black dotted lines. Finally, we have highlighted with blue solid circles, the points where black hole solutions arise and labeled them according to the corresponding solution.}
\label{fig:deg}
\end{figure}%
The roots of this fourth order polynomial are appropriately plotted in Fig.~\ref{fig:deg}. Moreover, since we are looking for solutions that are asymptotically Lifshitz, only strictly positive values of $\alpha$ must be considered. A careful analysis of Fig.~\ref{fig:deg} shows that four different real roots are found in the exponent intervals $z<1$ and $z>3$, while two complex conjugate solutions arise in the open interval $1<z<3$. Consequently, multiplicity of the roots appears in the transitions between these intervals at the exponents $z=1$ and $z=3$. Each of these cases will be treated separately. The plan is to start studying the linear space associated to the Euler equation (\ref{eq:elf}) in each case. After that, we address the fulfilling of the nonlinear constraint given by Eq. (\ref{eq:eln}), which defines a subspace of the previously defined linear space.

We will start our exhaustive analysis of the different
mentioned cases beginning with those for which the degeneracy
occurs, i.e.\ the exponents $z=1$ and $z=3$.

\subsection{Dynamical exponent $z=1$}

For a dynamical exponent $z=1$, the characteristic
polynomial (\ref{auxiliar}) reduces to
\begin{equation}
\alpha(\alpha-1)^2(\alpha-2)=0.
\end{equation}
Since we are only interested in asymptotically Lifshitz solutions the root $\alpha=0$ must be discarded. Additionally, because of the presence of a double root at $\alpha=1$ we need to supplement the two power-law linearly independent solutions with a logarithmic contribution. As a consequence the most general solution of the linear constraint is given by
\begin{equation}\label{eq:log1}
H(r)=\mu\frac{l^2}{r^2}+\frac{l}{r}\left[b+b_0\ln\left(\frac{r}{l}\right)\right].
\end{equation}
The substitution of (\ref{eq:log1}) into the nonlinear
constraint (\ref{eq:eln}) leads to
\begin{eqnarray}\label{eq:z1}
&&b_0\Bigl[(2{l}^2m^2+1)\frac{r}{l}-b_0\Bigr]\frac{r}{l}\ln\left(\frac{r}{l}\right)
-[4{l}^2m^2(1+l^2\lambda)+1]\frac{r^3}{l^3}\nonumber\\\nonumber\\
&&{}+(2{l}^2m^2+1)(b_0+b)\frac{r^2}{l^2}-\frac{b_0(4b-b_0)}4\frac{r}{l}-b_0\mu=0.
\end{eqnarray}
From this equation, it becomes clear that the conditions to be
satisfied are the following:
\begin{subequations}
\begin{eqnarray}
                      b_0&=&0,\\
4{l}^2m^2(1+l^2\lambda)+1&=&0,\\
             (2l^2m^2+1)b&=&0.
\end{eqnarray}
\end{subequations}
At this point we have two possible scenarios  depending on
whether the gravitational hair $b$ vanishes  or not. If we consider $b \neq 0$, then the
metric function $H$ and the cosmological constant read
\begin{equation}
H(r)=\mu\frac{l^2}{r^2} + b\frac{l}{r}, \qquad
\lambda=m^2=-\frac{1}{2l^2}.
\end{equation}
This solution corresponds to the $z=1$ black hole (\ref{eq:z=1bh})
found in \cite{Bergshoeff:2009aq,Oliva:2009ip}. On the
other hand, the case $b=0$ reduces to the static BTZ black hole
\cite{Banados:1992wn} but with the weaker constraint (\ref{eq:lambda2mz=1}).

We now turn to the other degenerate case for an exponent $z=3$.

\subsection{Dynamical exponent $z=3$}

In Fig.~\ref{fig:deg} another multiplicity can be appreciated at the
exponent $z=3$. In this case, the characteristic polynomial
(\ref{auxiliar}) becomes
\begin{equation}
(\alpha-2+\sqrt{5})(\alpha-2-\sqrt{5})(\alpha-2)^2=0,
\end{equation}
and thus the allowed strictly positive roots are $\alpha=2+\sqrt{5}$
and $\alpha=2$. Hence the general solution of the linear constraint in this case
is given by
\begin{equation}\label{eq:log3}
H(r)=\frac{l^2}{r^2}\left[M+M_0\ln\left(\frac{r}{l}\right)\right]
+M_1\left(\frac{l}{r}\right)^{\sqrt{5}+2}.
\end{equation}

The substitution of the metric function (\ref{eq:log3}) into the
nonlinear constraint (\ref{eq:eln}) together with the parameterizations (\ref{eq:ccz<>1}), which impose the Lifshitz behavior of the leading terms, leads to
\begin{eqnarray}
\label{eq:z3}
&&4M_0 \left[M_1\sqrt{5}(\sqrt{5}-1) \left(\frac{l}{r}\right)^{\sqrt{5}}-4M_0\right]
\left(\frac{l}{r}\right)^4\ln\left(\frac{r}{l}\right) \nonumber\\
&&{}+2\sqrt{5}(3-\sqrt{5}){M_1}^2\left(\frac{l}{r}\right)^{4+2\sqrt{5}} \nonumber\\
&&{}+\frac{\sqrt{5}(\sqrt{5}-1)}{10}\left[40M+\sqrt{5}(17-7\sqrt{5})M_0\right]{M_1}
\left(\frac{l}{r}\right)^{4+\sqrt{5}}\nonumber\\
&&{}+8\sqrt{5}(\sqrt{5}-2){M_1}
\left(\frac{l}{r}\right)^{2+\sqrt{5}}-M_0(16M+11M_0){\left(\frac{l}{r}\right)}^{4} \nonumber\\
&&{}+56M_0\left(\frac{l}{r}\right)^2=0.
\end{eqnarray}
It is clear from this expression that the only possibility is
to set $M_0=0=M_1$ while $M$ can be an arbitrary constant, yielding to the
solution
\begin{equation}
H(r)=M \frac{l^2}{r^2}, \qquad
\frac{\lambda}{13}=m^2=-\frac{1}{2l^2}.
\end{equation}
This latter is precisely the $z=3$ Lifshitz black hole (\ref{eq:z=3bh})
found in \cite{AyonBeato:2009nh}.

In what follows, we prove that for dynamical exponents
$z\not=1$ and $z\not=3$, Lifshitz black hole solutions of
NMG within the class of the Kerr-Schild Ansatz (\ref{eq:Lifansatz}) do not exist.

\subsection{Dynamical exponents $z<1$ and $z>3$\label{Subsec:nondeg}}

\begin{figure}[t]
\centering
\includegraphics[scale=1]{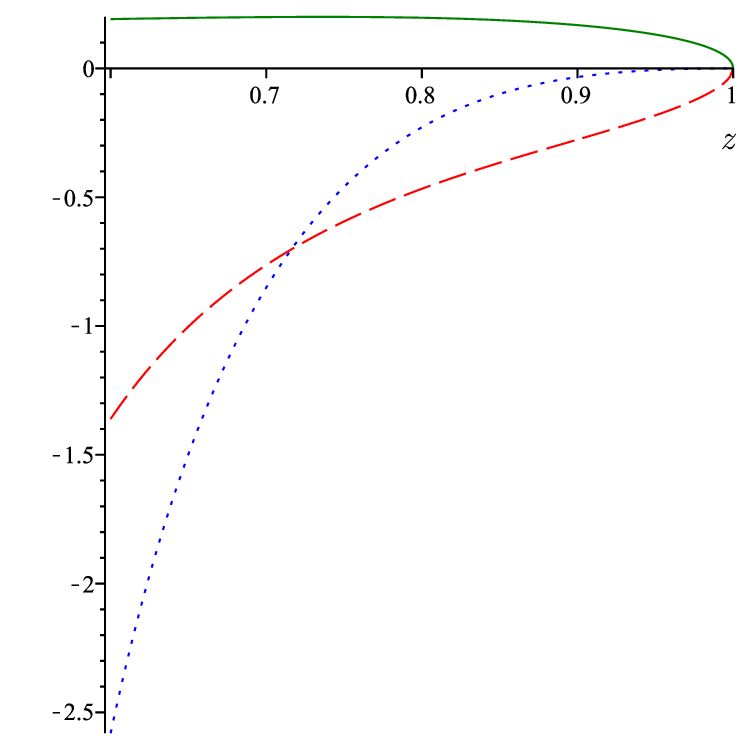}
\caption{The $z$-dependence in the interval $z<1$ of the coefficients in front of the powers exhibited in Eq.~(\ref{eq:nondeg}). The blue dotted line
represents the coefficient in front of $C_{++}
(l/r)^{\alpha_{++}}$, the green solid line
represents the coefficient in front of $C_{--}^{~2}
(l/r)^{2\alpha_{--}}$ and the red dashed line
represents the coefficient in front of $C_{+-}^{~2}
(l/r)^{2\alpha_{+-}}$.}
\label{fig:z1}
\end{figure}

\begin{figure}[t]
\centering
\includegraphics[scale=1]{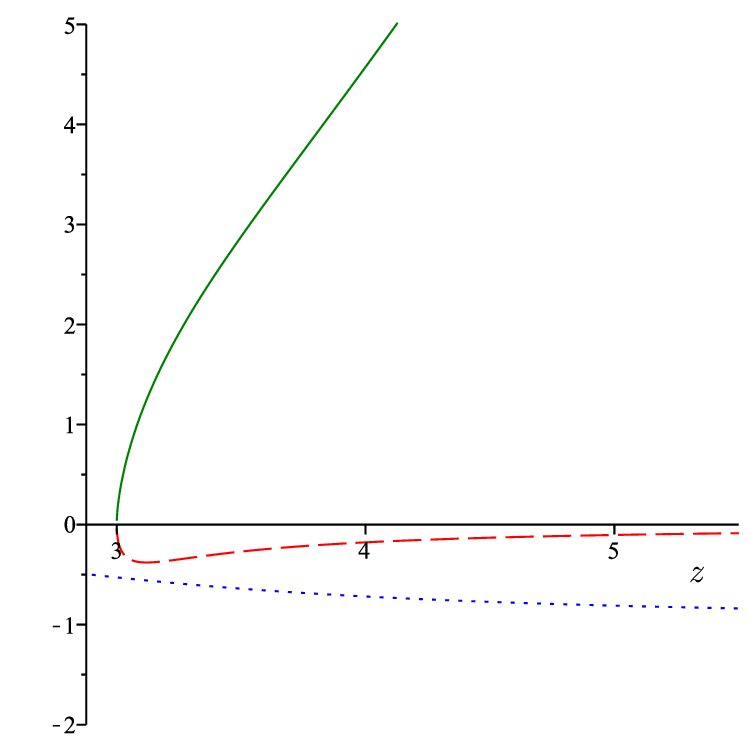}
\caption{The $z$-dependence in the interval $z>3$ of the coefficients in front of the powers exhibited in Eq.~(\ref{eq:nondeg}). The blue dotted line
represents the coefficient in front of $C_{++}
(l/r)^{\alpha_{++}}$, the green solid line
represents the coefficient in front of $C_{--}^{~2}
(l/r)^{2\alpha_{--}}$ and the red dashed line
represents the coefficient in front of $C_{+-}^{~2}
(l/r)^{2\alpha_{+-}}$.}
\label{fig:z3}
\end{figure}

The exponent intervals $z<1$ and $z>3$ can be analyzed together as they
share the same behavior regarding the nature of the solutions
of the characteristic polynomial as evidenced by Fig.~\ref{fig:deg}. Indeed, in this case, all of the roots of the
polynomial are real, and hence the solution is given by
\begin{eqnarray}\label{eq:Hnondeg}
H(r)&=&C_{++}\left(\frac{l}{r}\right)^{\alpha_{++}}
      +C_{+-}\left(\frac{l}{r}\right)^{\alpha_{+-}} \nonumber \\
  &&{}+C_{--}\left(\frac{l}{r}\right)^{\alpha_{--}},
\end{eqnarray}
where we have defined
\begin{subequations}
\begin{eqnarray}
\alpha_{\pm\pm}&=&\frac{z+1 \pm A_{\pm}}{2},\\
        A_{\pm}&=&\sqrt{5z^2-16z+13\pm2\sqrt{p}},\\
              p&=&4z^4-28z^3+73z^2-76z+28.
\end{eqnarray}
\end{subequations}
The reason why we have not included the power $\alpha_{-+}$ in (\ref{eq:Hnondeg}) is because it is not positive for $z<1$ nor for $z>3$ (see Fig.~\ref{fig:deg}); thus it does not comply with the asymptotically Lifshitz requirement. The substitution of (\ref{eq:Hnondeg}) and the parameterizations (\ref{eq:ccz<>1}) into the
nonlinear constraint (\ref{eq:eln}) gives an equation with the following general form
\begin{subequations}
\begin{eqnarray}\label{eq:nondeg}
&&\Biggl[
P_2(\alpha_{+-},z)C_{+-}^{~2}\left(\frac{l}{r}\right)^{2\alpha_{+-}}+
P_2(\alpha_{--},z)C_{--}^{~2}\left(\frac{l}{r}\right)^{2\alpha_{--}} \nonumber \\
&&{}+4P_1(\alpha_{++},z)C_{++}\left(\frac{l}{r}\right)^{\alpha_{++}} + \ldots \Biggr]/[8l^2(z^2-3z+1)]=0,\nonumber\\
\end{eqnarray}
where
\begin{eqnarray}
P_1(\alpha,z)&=&(\alpha-z-1)[\alpha^3-2z\alpha^2-(z^2-6z+4)\alpha\nonumber\\
&&{}+2z(z^2-3z+1)]\frac{}{},\\
P_2(\alpha,z)&=&(z-2)\alpha\left[2\alpha^2-(z-3)(2z-3)\right]\frac{}{}
\nonumber \\
&&{}-(4z^2-9z+1)\alpha^2\frac{}{}
\nonumber \\
&&{}+2(z^2-3z+1)(2z^2-3z+3),
\end{eqnarray}
\end{subequations}
and where the dots stand for a superposition of other powers of $l/r$ that will not be required to establish our conclusion. We have plotted in Fig.~\ref{fig:z1} and Fig.~\ref{fig:z3} the $z$-dependent part of the coefficients in front of the powers exhibited in Eq.~(\ref{eq:nondeg}), from which it is possible to realize that none of them vanish for $z<1$ and $z>3$, respectively. As a result, we may conclude that the only possibility is the vanishing of the constants $C_{++}$, $C_{+-}$ and $C_{--}$. In other words, no linear direction is compatible with the nonlinear constraint.%

\subsection{Dynamical exponents $1<z<3$\label{Subsec:complex}}

For the remaining case $1<z<3$, the characteristic polynomial
has two real roots $\alpha_{++}$ and $\alpha_{-+}$ and two
complex conjugate roots $\alpha_{+-}$ and $\alpha_{--}$ which in turn
implies that the general solution of the Euler equation reads
\begin{subequations}
\begin{eqnarray}\label{Hcomplex}
H(r)&=&  C_{++}\left(\frac{l}{r}\right)^{\alpha_{++}}
       + C_{-+}\left(\frac{l}{r}\right)^{\alpha_{-+}}\nonumber\\
&&{}+ \left(\frac{l}{r}\right)^{(z+1)/2} \Biggl[
     C_3\cos\left(\frac{A_c}{2}\ln\frac{r}{l}\right) \nonumber \\
&&{}-C_4\sin\left(\frac{A_c}{2}\ln\frac{r}{l}\right)  \Biggr],\\
\nonumber\\
A_c&=&\sqrt{-5z^2+16z-13+2\sqrt{p}}.
\end{eqnarray}
\end{subequations}

The substitution of the solution (\ref{Hcomplex}) to the linear equation in the interval $1<z<3$ and the parameterizations (\ref{eq:ccz<>1}) into the non-linear constraint (\ref{eq:eln}) has the following form after being separated into functionally independent expressions whose coefficients must vanish independently
\begin{subequations}
\begin{eqnarray}\label{eq:complex11}
&&\Biggl\{
     4P_3(\alpha_{++},z)C_{++}\left(\frac{l}{r}\right)^{\alpha_{++}}
-4P_3(\alpha_{-+},z)C_{-+}\left(\frac{l}{r}\right)^{\alpha_{-+}} \nonumber \\
&&{}+\left[[B_1(z)C_3+B_2(z)C_4]
\sin\left(\frac{A_c}{2}\ln\frac{r}{l}\right)\right.\nonumber \\
&&{}-\left.[B_2(z)C_3-B_1(z)C_4]
\cos\left(\frac{A_c}{2}\ln\frac{r}{l}\right)
\right] \left(\frac{l}{r}\right)^{(z+1)/2} \nonumber \\
&&{}+\dots\Biggr\} /[8l^2(z^2-3z+1)]=0,
\end{eqnarray}
where we have posited
\begin{eqnarray}
P_3(\alpha,z)&=&(z-1)\alpha^2(\alpha-2z+1)
-(z^3-4z^2+3z-2)\alpha\frac{}{}\nonumber\\
&&{}+2(z^2+1)(z^2-3z+1)\frac{}{},\\
B_1(z)&=&(2\bar{\alpha}_{+-}-z-1)[2(z-1)\bar{\alpha}_{+-}
(\bar{\alpha}_{+-}-z-1)\frac{}{}\nonumber\\
&&{}+5z^3-11z^2+3z-1],\frac{}{} \\
B_2(z)&=&2(z-1)(z-5)\bar{\alpha}_{+-}(\bar{\alpha}_{+-}-z-1)\frac{}{}\nonumber\\
&&{}+5z^4-20z^3+18z^2-24z+13,\frac{}{} \\
\bar{\alpha}_{+-}&=&\frac{z+1+A_c}{2}.
\end{eqnarray}
\end{subequations}%
\begin{figure}[t]
\centering
\includegraphics[scale=1]{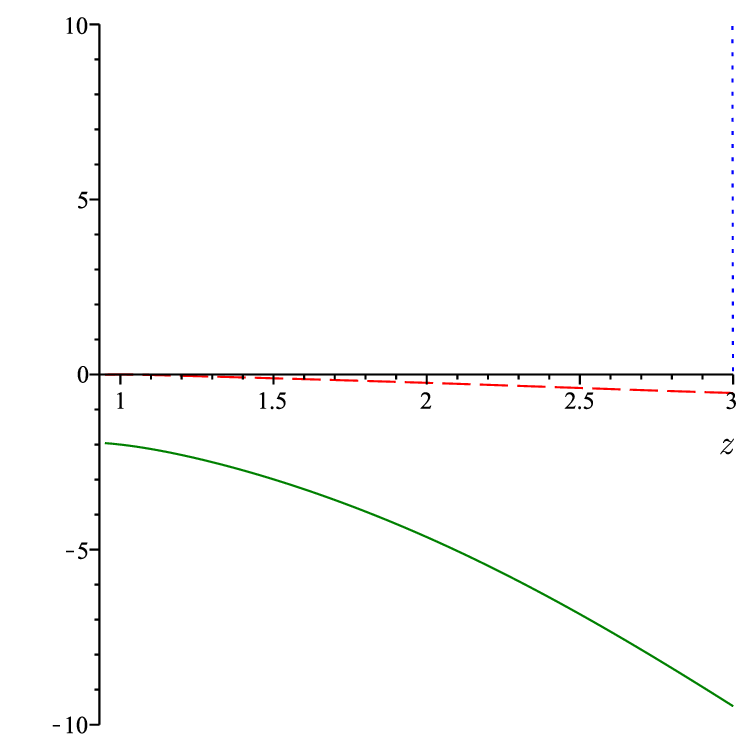}
\caption{The $z$-dependence in the interval $1<z<3$ of the coefficients in front of the functionally independent expressions exhibited in Eq.~(\ref{eq:complex11}).
The red dashed line represents the coefficient
in front of $C_{++}(l/r)^{\alpha_{++}}$, the
green solid line represents the coefficient
in front of $C_{-+} (l/r)^{\alpha_{-+}}$, the
blued dotted line represents the determinant
$B_1(z)^2+B_2(z)^2$ of the homogeneous linear system composed by
the constraints containing the constants $C_3$ and $C_4$.}
\label{fig:zcomplex}
\end{figure}%
Here, the dots stand for a superposition of other functionally independent functions of $r$ whose explicit forms will not be required to establish our conclusion. We have plotted in Fig.~\ref{fig:zcomplex} the coefficients in front of $C_{++}(l/r)^{\alpha_{++}}$ and $C_{-+}(l/r)^{\alpha_{-+}}$, they do not vanish in the interval $1<z<3$ and thus $C_{++}=0=C_{-+}$. Now, from Eq.~(\ref{eq:complex11}) we can see that the vanishing of the two coefficients accompanying the trigonometric functions give rise to a homogenous linear system for $C_3$ and $C_4$. The determinant of the matrix associated to the linear system is given by $B_1(z)^2+B_2(z)^2$ and is also represented in Fig.~\ref{fig:zcomplex}, it can be seen that it is nonvanishing in the interval $1<z<3$. This means the matrix is invertible and the linear system has only the trivial solution $C_3=0=C_4$.

As a result, from our analysis we can state that the only static and axisymmetric
asymptotically Lifshitz black hole solutions of NMG within the
Kerr-Schild ansatz (\ref{eq:Lifansatz}) are those previously known with dynamical
exponents $z=1$ and $z=3$.

\section{Corresponding Lifshitz solitons of new massive gravity}

It is well-known that starting from a static black hole solution in
three dimensions and operating a double Wick rotation, one can
generate a solution describing a gravitational soliton. As shown in \cite{Gonzalez:2011nz} in the context of Lifshitz background, a black hole with dynamical exponent $z$ is mapped into a soliton with dynamical exponent
$z^{-1}$. This is due to the fact that the two-dimensional Lifshitz algebras
with dynamical exponents $z$ and $z^{-1}$ are isomorphic.
There are two known solitons that have been generated by this mechanism in new massive gravity. The first one is the $z=1$ soliton found by Oliva \emph{et al.} \cite{Oliva:2009ip} which is given by
\begin{subequations}\label{eq:z=1sol}
\begin{equation}
ds^2=l^2[-(\tilde{b}+\cosh\rho)^2 d\tilde{t}^2 +d\rho^2 +\sinh^2\rho d\tilde{x}^2],
\end{equation}
where $\tilde{b}\equiv{b}/\sqrt{4\mu+b^2}$ and
\begin{equation}
\lambda=m^2=-\frac{1}{2l^2}.
\end{equation}
\end{subequations}
The case $b=0$ corresponds to the standard AdS$_3$ soliton which exists modulo the weaker constraint (\ref{eq:lambda2mz=1}). Under the coordinate transformation
\begin{equation}
(\tilde{t},\rho,\tilde{x})\mapsto
\!\Biggl(\!t=\frac{2\tilde{b}}{b}l\tilde{t},
r=\frac{lb}{2\tilde{b}}(\tilde{b}+\cosh\rho),
x=\frac{2\tilde{b}}{b}l\tilde{x}\!\Biggr)\!,\nonumber
\end{equation}
this metric can be rewritten in a way where the $z=1$ asymptotic scaling symmetry is manifest
\begin{eqnarray}
ds^2&=&-\frac{r^2}{l^2}dt^2
       +\frac{l^2}{r^2}\left(1-\mu\frac{l^2}{r^2}-b\frac{l}{r}\right)^{-1}dr^2 \nonumber \\
   &&{}+\frac{r^2}{l^2}\left(1-\mu\frac{l^2}{r^2}-b\frac{l}{r}\right)dx^2.
   \label{eq:z=1solK}
\end{eqnarray}

The other soliton of NMG is the $z=1/3$ Lifshitz soliton found by Gonz\'alez \emph{et al.} in \cite{Gonzalez:2011nz} and described by the metric
\begin{subequations}\label{eq:z=1/3sol}
\begin{equation}
ds^2=\tilde{l}^2(-\cosh^2\rho d\tilde{t}^2+d\rho^2+\cosh^4\rho\sinh^2\rho d\tilde{x}^2),
\end{equation}
for the coupling constants
\begin{equation}
\frac{\lambda}{13}=m^2=-\frac{1}{18l^2},
\end{equation}
\end{subequations}
where $l=\tilde{l}/3$. The use of the coordinates
\begin{eqnarray}
(\tilde{t},\rho,\tilde{x})&\mapsto&
\Bigl( t= \frac{{3l}\tilde{t}}{\sqrt{M}},
r=  M^{3/2}{l}\cosh^3\rho,
x =  \frac{3{l}\tilde{x}}{M^{3/2}}
\Bigr),\nonumber
\end{eqnarray}
reveals the exponent $z=1/3$,
\begin{eqnarray}
ds^2&=&-\frac{r^{2/3}}{l^{2/3}}dt^2
       +\frac{l^2}{r^2}\left(1-M\frac{l^{2/3}}{r^{2/3}}\right)^{-1}dr^2
        \nonumber \\
   &&{}+\frac{r^2}{l^2}\left(1-M\frac{l^{2/3}}{r^{2/3}}\right)dx^2.
   \label{eq:z=1/3solK}
\end{eqnarray}

A simple analysis shows that both solitons of new massive gravity cannot be written in the standard Kerr-Schild form for a real Lorentzian manifold.\footnote{Although complex generalizations of the Kerr-Schild ansatz are known \cite{PS}, we do not consider this possibility here.} However, both can be written as part of the following metric ansatz characterized by a single function like the Kerr-Schild one
\begin{equation}\label{eq:ansatzsoliton}
ds^2=-\frac{r^{2z}}{l^{2z}}dt^2 + \frac{l^2}{r^2}\frac{dr^2}{1-K(r)}+\frac{r^2}{l^2}[1-K(r)]dx^2.
\end{equation}
Inspired by the analogy with our previous results on black holes this single function ansatz will be our starting point to study soliton configurations. This time the Lifshitz boundary condition is $K(\infty)=0$.
As in the black hole case, in spite of the fact that the nontrivial
equations $E_{t}^{~t}=0$, $E_{r}^{~r}=0$ and $E_{x}^{~x}=0$ are
 linearly independent, only two of them are actually
differentially independent. More precisely, the combination
$E_{r}^{~r}-E_{x}^{~x}=0$ will act now as the linear constraint
while the equation $E_{x}^{~x}=0$ will yield to the nonlinear
one. Indeed, we have
\begin{widetext}
\begin{subequations}
\begin{eqnarray}
\frac{4m^2l^4}{1-K}\left(E_{r}^{~r}-E_{x}^{~x}\right)&=&
{r}^{4}K'''' +2 {r}^{3} \left( 4+z \right) K''' -{r}^2 \left( 5 z+3 \right)  \left( z-4 \right)K''-2 zr \left( 2+z \right)  \left( 3 z-5 \right) K'\nonumber\\
&&{}+2 z \left(z- 1\right)  [ ({
z}^2   -3 z   +1)K -({z}^2-3 z+1+2 {l}^2m^2 )]=0, \label{eq:LinearSoliton} \\\nonumber\\
-16l^4m^2E_{x}^{~x}&=&
4 (1-K) r^4 K''''-2[ r K'-2 (z+9) (1-K) ]r^3 K'''+(r^2K'')^2\frac{}{}\nonumber\\
&&{}
-2[ r (z+2) K'+2(4 z^2-11 z-17) (1-K) ]r^2 K''
+4 (2 z^2-4 z+1) (rK')^2 \frac{}{}\nonumber \\
&&{}
-4  [(3 z^3+7 z^2-21 z-2) (1-K)-2z l^2  m^2] r K'\frac{}{}\nonumber \\
&&{}-4 (z^2-3 z+1) (z^2-z-1) (1-K)^2 -16m^2l^2[  z^2  (1-K)+ l^2 \lambda]
=0, \label{eq:NLCSoliton}
\end{eqnarray}
\end{subequations}
\end{widetext}
where (\ref{eq:LinearSoliton}) corresponds again to a fourth-order Euler differential equation and (\ref{eq:NLCSoliton}) plays the role of the nonlinear constraint. Since pure Lifshitz spacetimes must be solution of the field equations at least asymptotically, the constraints (\ref{eq:lambda}) and (\ref{eq:mp}) still hold, and thus the equation (\ref{eq:LinearSoliton}) becomes an homogeneous equation. The corresponding characteristic polynomial this time is%
\begin{figure}[b]
\centering
\includegraphics[scale=1]{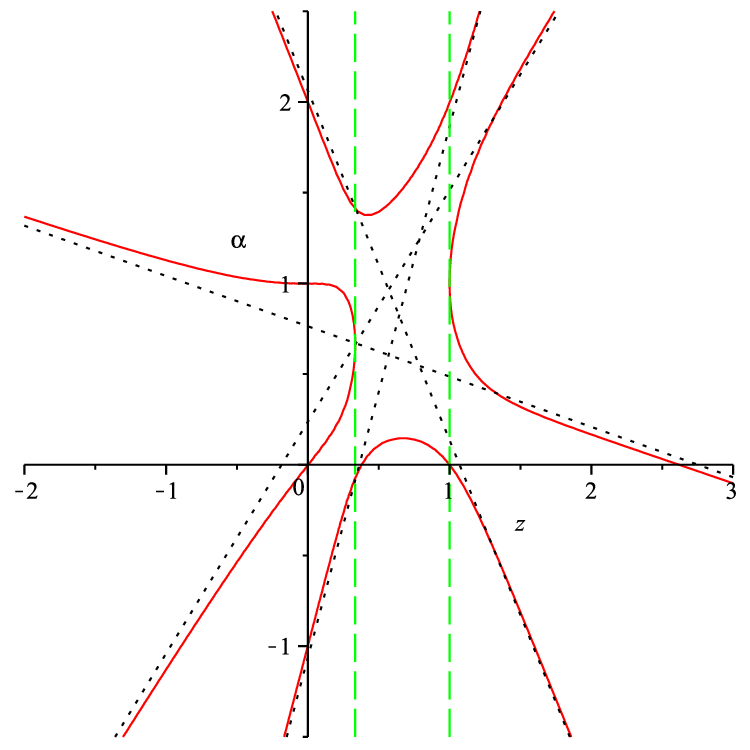}
\caption{Real solutions to the characteristic polynomial
equation associated to the soliton's Euler equation. It is to
be noted that while most values of $z$ correspond to 4 values
of $\alpha$, multiplicity arises in $z=1/3$ and $z=1$ as roots
$\alpha=2/3$ and $\alpha=1$ respectively are degenerate. Also,
in the interval $1/3<z<1$, two of the roots are actually
complex (for which they do not appear in the graph) and only
the two real solutions appear in the figure.}
\label{fig:degsolitones}
\end{figure}%
\begin{eqnarray}\label{eq:alphaSoliton}
\alpha^4-2 \left( z+1 \right) {\alpha}^{3}- \left(5
z^2 - 11 z+1 \right) {\alpha}^2 &&{} \nonumber \\
+ \left( 1+z \right)  \left( 6 {z}^2-9
z+2 \right) \alpha
 &&{} \nonumber \\
+2 z \left( z-1 \right)  \left( {z}^2-3 z+1
 \right)&=&0,
\end{eqnarray}
and exhibits multiplicity for $z=1$ and $z=1/3$ as shown in
Fig.~\ref{fig:degsolitones}.

We would like to stress that performing the changes
$z_{\mathrm{s}} = 1/z_{\mathrm{bh}}$ and  $\alpha_{\mathrm{s}}
= \alpha_{\mathrm{bh}}/z_{\mathrm{bh}} $ on the characteristic
polynomial associated to the solitonic case
(\ref{eq:alphaSoliton}), it reproduces the characteristic
polynomial obtained for the black hole (\ref{auxiliar}). This
suggests that there may exist a map between the solitonic
linear and nonlinear constraints and those associated to the
black hole solutions. In fact, performing the following
redefinitions\footnote{We thank R.~Troncoso for encouraging us to explore the relations between the black holes and the solitons in order to prove a similar uniqueness result also in the last case.}
\begin{eqnarray}
z_{\mathrm{bh}} &=& \frac{1}{z_{\mathrm{s}}},\qquad\qquad\qquad
l_{\mathrm{bh}} = \frac{l_{\mathrm{s}}}{z_{\mathrm{s}}},\nonumber\\
\frac{r_{\mathrm{bh}}}{l_{\mathrm{bh}}} &=&
\left(\frac{r_{\mathrm{s}}}{l_{\mathrm{s}}}\right)^{z_{\mathrm{s}}},\qquad
H(r_{\mathrm{bh}}) = K(r_{\mathrm{s}}),
\label{map}
\end{eqnarray}
we can relate these two pairs of constraints (the subindexes
``s'' and ``bh'' standing for solitonic and black hole,
respectively). As a result, it is possible to find the
corresponding Lifshitz solitons via this map. Moreover, since
the only Lifshitz black holes found within the Kerr-Schild
ansatz are those with dynamical exponents $z=1$ and $z=3$, we can state as a consequence that the only soliton solutions within our working single-function ansatz are those previously known with $z=1$ and $z=1/3$. Just to
be complete, we mention that a similar exhaustive analysis to
the one done in the black hole case will yield to the same
conclusion.

\subsection{The $z=1$ and $z=1/3$ solitons}

As it can be seen from (\ref{map}), the value $z=1$ corresponds
to a fixed point of the map, and hence the metric function does
not change. More precisely, we obtain
\begin{eqnarray}
K(r) = \mu\frac{l^2}{r^2} +b\frac{l}{r},\quad
\lambda=m^2=-\frac{1}{2l^2},
\end{eqnarray}
while for $b=0$ we recover the AdS soliton which requires the weaker constraint (\ref{eq:lambda2mz=1}). These $z=1$ solitons correspond to those
rewritten in (\ref{eq:z=1solK}) and found in \cite{Oliva:2009ip}.

Using the map (\ref{map}) the black hole found for
$z_{\mathrm{bh}}=3$ will correspond to a soliton with
$z_{\mathrm{s}}=1/3$. The soliton solution would then be
\begin{equation}
K(r)= M\frac{l^{2/3}}{r^{2/3}}, \quad \frac{\lambda}{13}=m^2=-\frac{1}{18l^2}.
\end{equation}
This corresponds to the $z=1/3$ Lifshitz soliton solution of NMG as described in (\ref{eq:z=1/3solK}) and
found in \cite{Gonzalez:2011nz}.

\section{Further works and conclusions}

We have established that the only static, axisymmetric Lifshitz black holes that can be obtained from the pure Lifshitz spacetimes through a Kerr-Schild transformation are those previously found in the literature with dynamical exponents $z=1$ and $z=3$. Usually, the Kerr-Schild transformation is an interesting tool for standard general relativity since it yields to a linear differential equation. In our case, we have shown that in spite of considering NMG which is of higher-order, the Kerr-Schild transformation on NMG yields to an Euler differential equation in addition to a nonlinear constraint. The strategy we follow was simple and it consists on first characterizing the linear space of solutions of the Euler equation for each dynamical exponent and then exploring the compatibility of this linear space with the nonlinear constraint.

Furthermore, we have been able to map these black hole constraints to
their corresponding solitonic ones and proved in this way that the soliton counterparts with dynamical exponents $z=1$ and $z=1/3$ are the only Lifshitz
solitons that can exist in NMG within certain ansatz depending on a single function. We have put in light a relation between the existence of black hole solutions with the multiplicity of the roots of the characteristic polynomial associated to the Euler equation. This feature appears as well in the case of solitons where multiplicity arises for $z=1/3$ and $z=1$. This partially answers what is special about this exponents in NMG. An interesting work will then consist in giving a more deep physical and/or mathematical explanation for this curiosity. In Ref.~\cite{AyonBeato:2010tm}, the authors have found the analogous solutions of those of new massive gravity for general quadratic corrections of the Einstein gravity in arbitrary dimension. We believe that the arguments established here can be generalized to the case of these higher-dimensional Lifshitz black hole solutions. There also exist examples for which the Kerr-Schild transformation allows to obtain interesting solutions in presence of matter sources as for example in the context of eleven-dimensional supergravity with the $3-$form \cite{Dereli:1986cm}. Hence, it will be interesting to explore whether the mechanism presented here can also be useful to obtain Lifshitz black hole solutions in presence of matter source as for example those derived recently in the context of nonlinear electrodynamics \cite{Alvarez:2014pra} or in presence of scalar fields \cite{Bravo-Gaete:2013dca,Correa:2014ika}.

\begin{acknowledgments}
The authors thank N.~Breton, A.~Garcia and R.~Troncoso for useful discussions. This work has been partially supported by grants 1130423,
11090281 and 1121031 from FONDECYT, by grants 175993 and 178346
from CONACyT and by CONICYT, Departamento de Relaciones
Internacionales ``Programa Regional MATHAMSUD 13 MATH-05.''
M.M.J-A is supported by ``Programa de Becas Mixtas'' from CONACyT and
``Plataforma de Movilidad Estudiantil Alianza del Pac\'{\i}fico.''
E.A-B is supported by ``Programa de Estancias Sab\'aticas al Extranjero''
from CONACyT and ``Programa Atracci\'{o}n de Capital Humano
Avanzado del Extranjero, MEC'' from CONICYT.
\end{acknowledgments}


\end{document}